# Spin-lattice coupling in an epitaxial $NdFeO_3$ thin film

M.A. Khaled[1], J. Ruvalcaba[2], T. Fraga Córdova[2], M. El Marssi[1], H. Bouyanfif[1]

[1] *LPMC UR2081, Université de Picardie Jules Verne, 33 Rue Saint Leu, 80000 Amiens, France*
[2] *División de Ciencias e Ingenierías, Universidad de Guanajuato campus León, México*

**Abstract**

Rare-earth orthoferrite $RFeO_3$ materials such as $NdFeO_3$ are strongly studied because of their fascinating magnetic properties and their potential applications. Here, we show the successful epitaxial synthesis of parasitic-free NFO thin film by pulsed laser deposition on (001)-$SrTiO_3$. High-resolution X-ray diffraction shows a coherent growth and a tetragonal-like structure of a tensile strained 80 nm thick NFO film in contrast with the bulk orthorhombic state. Room temperature magnetometry indicates a bulk-like antiferromagnetic state for the NFO film. Temperature-dependent X-ray diffraction and magnetometry highlight a significant spin-lattice coupling at the Néel Temperature while a new magneto-structural instability is discovered around 250°C that needs further investigation.

**Keywords**

Antiferromagnetic, Thin films, Epitaxial strain, Structure.

**Introduction**

Orthoferrites $RFeO_3$ are actively investigated due to their promising properties in spintronic and ultrafast applications [1-3]. Being antiferromagnetic, orthoferrites are robust against magnetic perturbations, produce no stray fields and their spin dynamics are typically two orders of magnitude faster compared to ferromagnets [4-7]. Particularly, Neodymium ferrite $NdFeO_3$ (NFO) adopts a perovskite $ABO_3$ structure with the A and B sites occupied by rare earth $Nd^{3+}$ and $Fe^{3+}$ ions, respectively. The crystallographic structure is described by an orthorhombic Pbnm symmetry with an $a^-a^-c^+$ oxygen octahedra rotation/tilt system. While a non-colinear antiferromagnetic spin ordering for $Fe^{3+}$ is observed at Néel temperature $T_N(Fe^{3+}) \simeq 760K$, the Neodymium ions typically show an antiferromagnetic ordering at very low temperature [8]. Bulk NFO shows a spin reorientation transition (SRT) of the $Fe^{3+}$ sublattice in a temperature range of 100K to 170K. Exchange and Dzyaloshinskii-Moriya interactions within and between the different sublattices ($Nd^{3+}$-$Nd^{3+}$, $Fe^{3+}$-$Fe^{3+}$, $Nd^{3+}$-$Fe^{3+}$) explain these varieties of magnetic structures and instabilities. For obvious reasons of miniaturization but also to understand size and strain effects, epitaxial thin films of such a material are required. For instance, NFO epitaxial thin films are still unexplored and strain effect on structural and magnetic properties remains obscure. A few reports of other $RFeO_3$ thin films can be found in the literature and are focused on low-temperature properties. Notably, $DyFeO_3$ thin films were investigated and a strong strain effect on $T_{SRT}$ has been evidenced while bulk-like properties were detected on $TmFeO_3$ [5,9]. The aim of this work is therefore to investigate coherently grown NFO thin film on a non-common measurement range exploring the possible impact of strain on the $Fe^{3+}$ antiferromagnetic to paramagnetic transition.

## Experimental details

The NFO thin film was grown on (001)-$SrTiO_3$ using pulsed laser deposition equipped with an excimer KrF laser (248 nm wavelength and 4 Hz pulse frequency). A homemade NFO ceramic target was used for the thin film growth. Substrate temperature and Oxygen pressure during growth were respectively fixed at 740°C and 0,1 mBar. Target-substrate distance and Laser fluency are respectively 4.5 cm and 1.7 J/cm$^2$. High-resolution X-Ray diffraction characterizations ($\theta$-2$\theta$, rocking curve, reflectivity, reciprocal space mapping) were performed using a D8 Brucker diffractometer ($\lambda$ = 1.54056 Å). A Quantum Design Cryogen-free system equipped with a Vibrating Sample Magnetometer allowed us to acquire M-H hysteresis loops from room temperature up to 700°C.

## Results and discussions

Figure 1 presents the room temperature X-ray diffraction characterization of the NFO film. The thin film is fully epitaxial with *(00l)* pseudo-cubic (pc) orientation along the growth direction and free of parasitic phases (see figure 1a). X-ray reflectivity scan was also collected (insert figure 1a) showing a large number of oscillations as proof of the flat nature and high quality of the interfaces. The simulation of the reflectivity enables us to verify the thickness of the NFO film (80 nm). The pseudo-cubic out-of-plane lattice parameter deduced from the Bragg law and the diffraction peak position in the $\theta$-2$\theta$ scan is 3.956 Å. This value is very close to the bulk pseudo-cubic $b_{pc}$~$b_O/\sqrt{2}$ value ($c_{pc}$~$c_O/2$ = 3.881 Å, $a_{pc}$~$a_O/\sqrt{2}$ = 3.855 Å and $b_{pc}$~$b_O/\sqrt{2}$ = 3.952 Å for bulk NFO) [12,13]. This suggests that the NFO thin film growth is along $[110]_O$ orthorhombic (parallel to [001] direction of the STO substrate). The similar value of the thin film out-of-plane lattice parameter, when compared to its bulk counterpart, suggests the absence of any tensile strain effect. Reciprocal space mapping around the (103) reflection of the substrate shows a coherent growth of our thin film (figure 1b). The NFO reflection is indeed perfectly aligned with the STO's in the $q_x$ direction indicating a similar in-plane lattice parameter (3.905 Å). The same result is obtained for the orthogonal in-plane direction $((013)_{pc}$ reflection, not shown) indicating square-like basis and tetragonal-like structure.

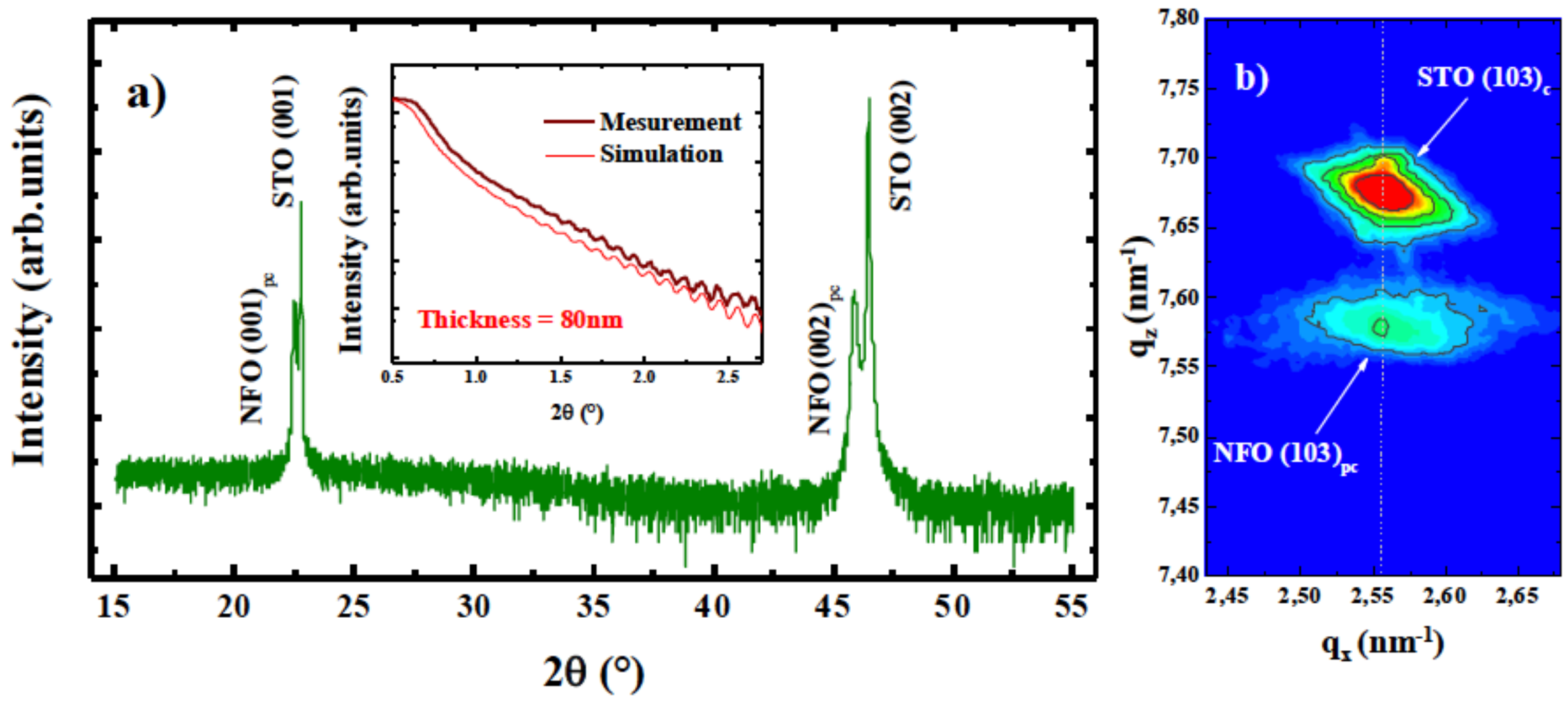

Figure 1. (a) θ-2θ scan of the NFO film deposited on (001)-STO substrate. Insert: X-ray reflectivity and simulation of the NFO thin film. (b) Reciprocal space mapping around the (103) reflection of the STO substrate.

To investigate the high-temperature structural behavior, θ-2θ scans were collected at different temperatures up to 700°C. From each scan, an out-of-plane lattice parameter is calculated using the Bragg law and the results are presented in Figure 2. To rule out any substrate effect the STO lattice parameters over the whole range of temperature are also shown (see insert figure 2a). An overall increase of the lattice parameters is observed both for the NFO thin film and the STO substrate (linear thermal dilatation). While the STO shows a perfect linear dilatation, the NFO thin film presents two anomalies on the lattice parameter evolution with temperature. On cooling from high temperature, a change of slope is evidenced close to the bulk $T_N$ highlighting a coupling between the magnetic and lattice degrees of freedom. On further cooling down a second anomaly at $T_1$ ~200 °C is observed both on the lattice parameter and the rocking curve (figure 2a and 2b). Moreover, the Full Width at Half Maximum (FWHM) of the rocking curve shown in figure 2c exhibits a plateau below this second structural anomaly observed at $T_1$. The only known other instability in NFO bulk is the spin reorientation transition occurring between 100K and 170K [10]. This second anomaly may belong to the antiferromagnetic transition starting at $T_N$. The limited resolution of our in-house diffractometer unfortunately does not allow us to pin down the exact symmetry at high temperatures. We would like to precise again that no anomalies are detected on the STO substrate showing that detected anomalies are therefore intrinsic to the NFO thin film.

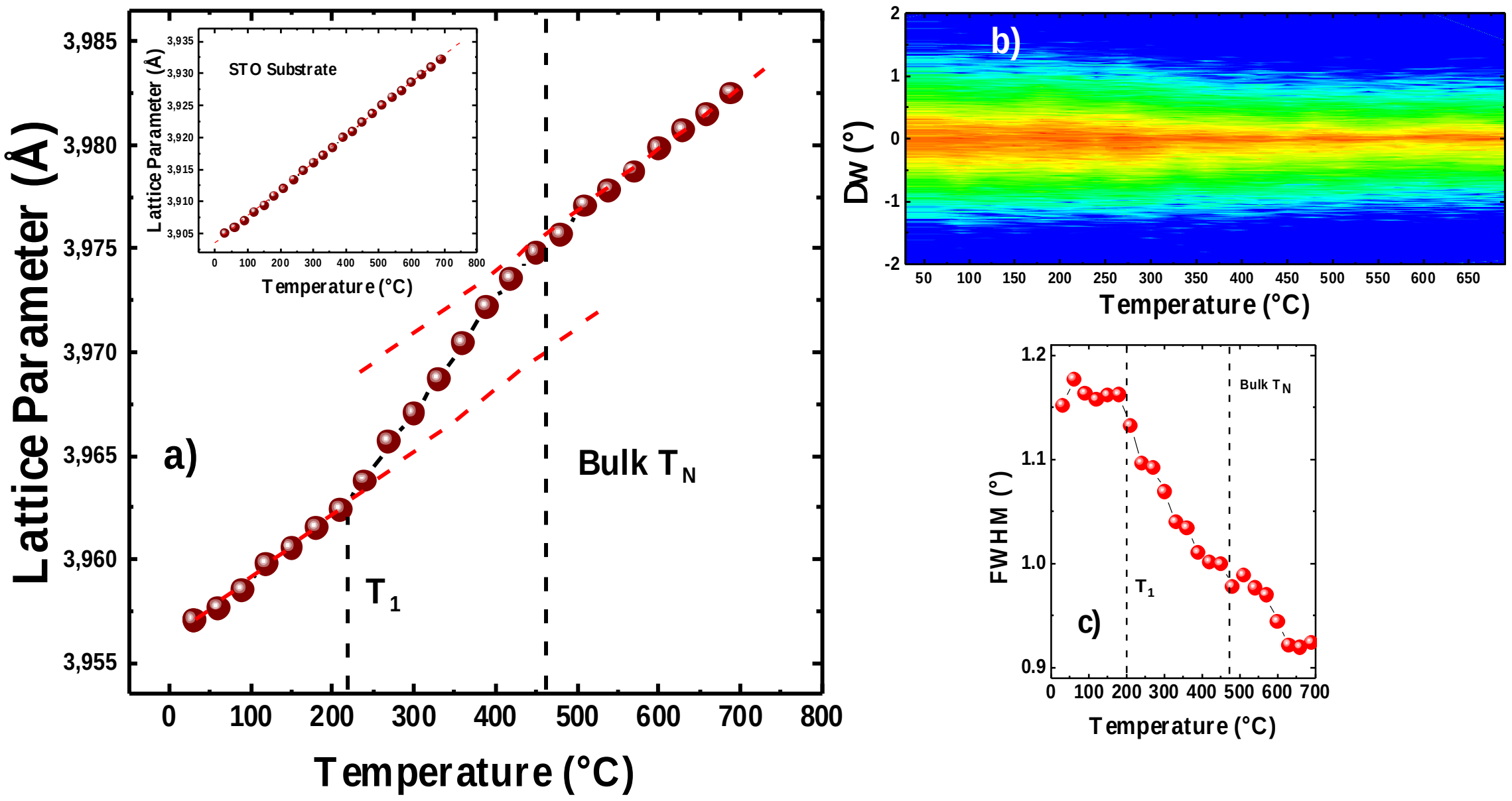

Figure 2. (a) Evolution with the temperature of the NFO out-of-plane lattice parameter (insert: STO lattice parameter versus temperature). The dashed black lines indicate the Néel transition $T_N$ and the anomaly's temperature $T_1$. The two dashed red lines are guides for the eyes. (b) Evolution with the temperature of rocking curves performed on the $(002)_{pc}$ NFO diffraction peak and the corresponding (c) Full Width at Half Maximum. Two dashed vertical lines indicate positions of anomalies.

To evidence the correlation of structural change with spin degrees of freedom, we have collected M(H) hysteresis loops from room temperature up to 700°C. M(H) loops were collected with the magnetic field applied in the plane (hysteresis loops measured along the orthogonal in-plane direction do not show differences). The diamagnetic contribution from the substrate has been corrected and the results are shown in Figure 3.

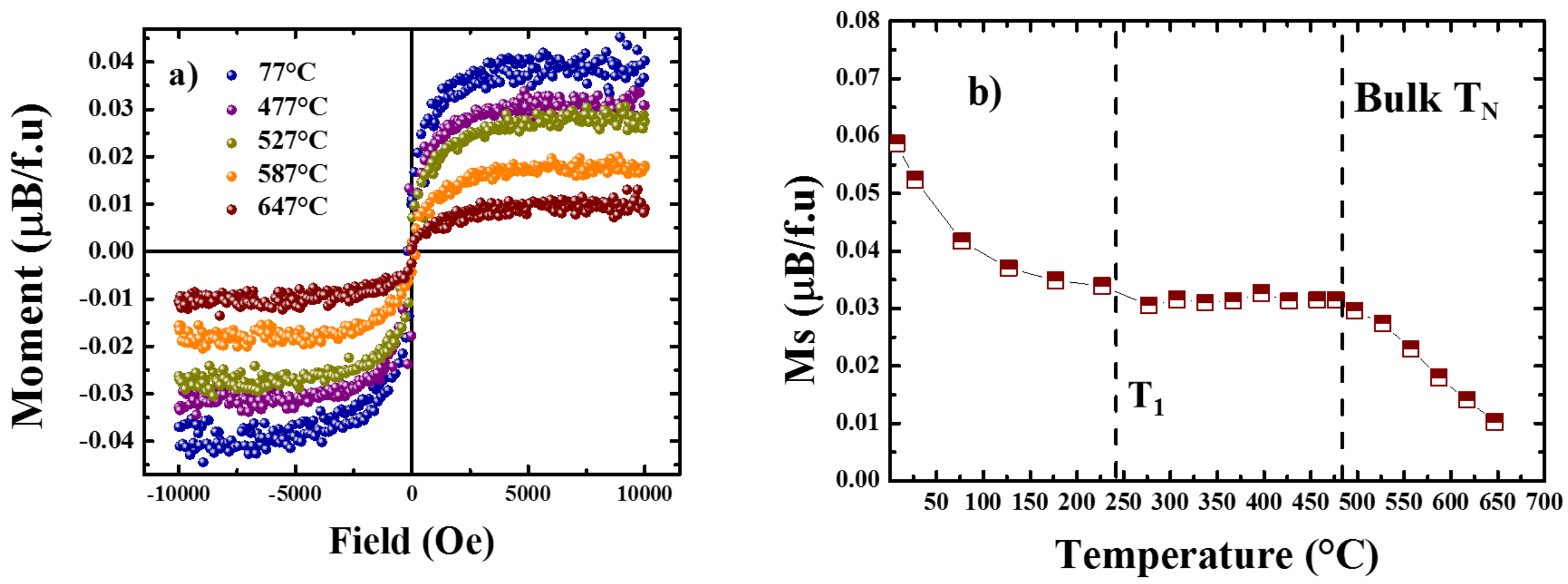

Figure 3. (a) M(H) hysteresis loops recorded at different temperatures. (b) Magnetic moment at saturation Ms versus temperature. The vertical dashed black lines indicate the two anomalies in Ms observed at $T_1$ and $T_N$.

The low-temperature hysteresis loops are slim in agreement with a canted antiferromagnetic state typical of orthoferrites as shown in figure 3a. To better understand the magnetic evolution on heating, the sample Magnetic moments at saturation (Ms) were estimated from the M(H) loops and are plotted in figure 3b. The room-temperature value is found to be Ms= 0.052 µB/f.u, slightly higher than the expected bulk value of ≃0.04 µB/f.u [10]. Similar to the lattice anomalies observed in figure 2 we also detect two anomalies on the Ms values at very close temperatures (at 200-250°C and 460°C). The high-temperature anomaly is attributed to the $Fe^{3+}$ antiferromagnetic transition to a paramagnetic state ($T_N$). Indeed, above $T_N$ (467°C in bulk) the weak ferromagnetic moment caused by the Dzyaloshinski-Moriya interaction is suppressed and a concomitant strong decrease of the magnetic moment at saturation is observed on the M(H) loops. Spin-lattice coupling is therefore evidenced from the combination of lattice parameters and Ms anomalies evidenced at the same $T_N$ temperature. The second lattice anomaly detected at $T_1$ (≃200-250°C) is also observed on the Ms behavior. Therefore, a spin-lattice coupling is also at stage that may be due to a structural transition. To identify the exact origin of this transition, diffraction using synchrotron sources is required (RSM thin film signals are weak). Analysis such as x-ray dichroism and absorption may also help understand the exact spin arrangement at such temperatures within nanometric thin.

**Summary**

In summary, parasitic-free and epitaxial NFO thin film was grown by pulsed laser deposition and characterized by high-resolution X-ray diffraction and vibrating sample magnetometry. In contrast to the bulk, 80nm thick antiferromagnetic NFO film is shown to be tetragonal-like due to tensile strain imposed by the STO cubic substrate. Temperature-dependent X-ray diffraction and magnetometry indicate two structural changes with significant spin-lattice coupling. While the $T_N$ is detected at a similar value compared to the bulk, an unusual anomaly is evidenced at $T_1$ (200°C-250°C). Additional structural and magnetic investigations are clearly required to find out the exact origin of such behavior prior to any NFO thin film applications.

**References**

[1]W.C.Koehler, E.O.Wollan, M.K.Wilkinson, Phys. Rev. **118**, 58 (1960).

[2]D.Treves, J. Appl. Phys. **36**, 1033 (1965).

[3]R.L.White, J. Appl. Phys. **40**, 1061 (1969).

[4]S.Schlauderer, C.Lange, S.Baierl, T.Ebnet, C.P.Schmid, D.C.Valovcin, A.K.Zvezdin, A.V.Kimel, R.V.Mikhaylovskiy, R.Huber, Nature **569**, 383 (2019).

[5]S.Becker, A.Ross, R.Lebrun, L.Baldrati, S.Ding, F.Schreiber, F.Maccherozzi, D.Backes, M.Kläui, G.Jakob, Phys. Rev. B **103**,024423 (2021).

[6]Y.Tokunaga, N.Furukawa, H.Sakai, Y.Taguchi, T.Arima, Y.Tokura, Nature Mater. **8**, 558 (2009).

[7]Y.Tokunaga, S.Iguchi, T.Arima, Y.Tokura, Phys. Rev. Lett. **101**, 097205 (2008).

[8]E.Bousquet, A.Cano, Physical Sciences Reviews, (2021).

[9]T.Y.Khim, M.J.Eom, J.S.Kim, B.G.Park, J.Y.Kim, J.H.Park, Appl. Phys. Lett. **99**, 072501 (2011).

[10]S.J.Yuan, W.Ren, F.Hong, Y.B.Wang, J.C.Zhang, L.Bellaiche, S.X.Cao, G.Cao, Phys. Rev. B **87**, 184405 (2013).

[11]W.Sławinski, R.Przeniosło, I.Sosnowska, E.Suard, J. Phys.: Condens. Matter **17** 4605-4614 (2005).

[12]M.Marezio, J.P.Remeika, P.D.Dernier, Acta Crystallogr. B **26**, 2008 (1970).

[13]V.A.Streltsov, N.Ishizawa, Acta Crystallogr. B **55,** 1 (1999).

[14]K.J.Choi, M.Biegalski, Y.L.Li, A.Sharan, J.Schubert, R.Uecker, P.Reiche, Y. B.Chen, X.Q.Pan, V.Gopalan, L.-Q.Chen, D.G.Schlom, C.B.Eom, Science 306, 1005 (2004).